\documentstyle[pra,aps,multicol]{revtex}

\newcommand{\be}{\begin{equation}}
\newcommand{\ee}{\end{equation}}

\begin{document}
\draft
\title{Unitary relation between a harmonic oscillator of time-dependent frequency 
and a simple harmonic oscillator with and without an inverse-square potential}
\author{Dae-Yup Song\footnote[1]{Electronic address: 
        dsong@sunchon.ac.kr}}
\address{ Department of Physics, Sunchon National University, Sunchon 
540-742, Korea}
\date{\today}
\maketitle
\begin{abstract}
The unitary operator which transforms a harmonic oscillator system of time-dependent 
frequency into that of a simple harmonic oscillator of different time-scale is found, 
with and without an inverse-square potential. It is shown 
that for both cases, this operator can be used in finding complete sets of wave 
functions of a generalized harmonic oscillator system from the well-known sets of the 
simple harmonic oscillator. Exact invariants of the time-dependent 
systems can also be obtained from the constant Hamiltonians of unit mass and frequency
by making use of this unitary transformation. 
The geometric phases for the wave functions of a generalized 
harmonic oscillator with an inverse-square potential are given.
\end{abstract}

\begin{multicols}{2}
It is certainly of importance to find a complete set of wave functions 
for a system of the time-dependent Hamiltonian [1-17]. 
It has long been known that a harmonic oscillator of time-dependent frequency 
with or without an inverse-square potential is the system of practical 
applications (see e.g. Ref. \cite{Brown}), where the wave functions are described 
in terms of solutions of classical equation of motion of the oscillator
without the inverse-square potential \cite{Lewis,LR,RR,KL,Ped}. 
In Ref. \cite{Song} it has been shown that, for a generalized harmonic oscillator 
system, the kernel of the system is determined by
the classical action. This is one of the basic reasons of the fact that the wave 
functions are described by the classical solutions. 
On the other hand, it has long been noticed that there exist 
classical canonical transformations which relate the (driven) harmonic oscillators 
of different parameters (see e.g. Refs. \cite{Ped,LL}). 
Recently, in Ref. \cite{BFL}, it has been shown that a driven harmonic oscillator of 
time-dependent frequency is related, through canonical transformations, to the simple 
harmonic oscillator of unit mass and unit frequency but with a different time-scale 
\cite{HR,LL}. This fact has 
been used to find the wave functions of a driven system which exactly agree with 
the known results \cite{Song,SongU}.

In this Rapid Communication, we will show that for both oscillators with and 
without an inverse-square potential, there is a unitary operator which transforms 
the harmonic oscillator systems of time-dependent frequency into those of the 
unit-mass and unit-frequency oscillators with different time-scales. 
This unitary operator can be used to find complete sets of wave functions of the 
systems with time-dependent parameters from the well-known sets of wave functions 
of the simple harmonic oscillator with \cite{Calogero} or without an inverse-square 
potential. It has been known that \cite{Lewis,LR,RR,KL,Ped} there exist {\em exact} 
invariants in the systems of time-dependent parameters which have 
long been used to 
find the wave functions \cite{Lewis,LR,RR,KL,Ped,Um,Ji,Ped2}.
As might have been implied by the classical treatments through canonical 
transformations, it will also be shown that, the {\em exact} quantum-mechanical 
invariants in oscillator systems of {\em time-dependent} parameters can be obtained 
from the {\em constant} Hamiltonians of unit mass and frequency (which, certainly, 
are invariants in their systems respectively), through the unitary transformation 
given here and those in Refs. \cite{Song,SongU}.  

The unit-mass harmonic oscillator of time-dependent frequency $w_0(t)$ is described 
by the Hamiltonian
\be
H_0(x,p,t)={p^2 \over 2}+{w_0^2(t)\over 2}x^2,
\ee
with the classical equation of motion
\be
\ddot{x} + w_0^2(t) x=0.
\ee
If we denote the two linearly independent solutions of Eq. (2) as $u_0(t)$ and $v_0(t)$,
the $\rho_0(t)$ defined by $\rho_0(t)=\sqrt{u_0^2+v_0^2}$ should satisfy
\be
{d^2 \over dt^2}{\rho_0}+w_0^2(t)\rho_0 - {\Omega_0^2 \over \rho_0^3}=0,
\ee
with a time-constant $\Omega_0$ ($\equiv\dot{v}_0u_0-\dot{u}_0v_0$). Without losing 
generality we assume that $\Omega_0$ is positive.
The wave functions $\psi_n^0(x,t)$ of the system should satisfy the
Schr\"{o}dinger equation
\be
O_0(t)\psi_n^0(x,t) =0, 
\ee
where $O_0(t)=-ih{\partial \over \partial t}+ H_0(x,p,t)$.
For the simple harmonic oscillator system of unit mass and frequency whose time is 
$\tau$, the wave functions $\psi_n^s(x,\tau)$ should satisfy
\be
O_s(\tau)\psi_n^s(x,\tau) =0,
\ee
where $O_s(\tau)=-ih{\partial \over \partial \tau}+ H^s$ with 
$H^s={1\over 2}(p^2+x^2).$
If $t$ and $\tau$ is related through the relation
\be
d\tau={\Omega_0 \over \rho_0^2} dt,
\ee
by defining the unitary operator 
\be
U_{w0}(\rho_0,\Omega_0)=\exp({i\dot{\rho_0} \over 2\hbar \rho_0}x^2)
  \exp[-{i\over 2\hbar}(\ln{\rho_0\over \sqrt{\Omega_0}})(xp+px)],
\ee
one may find the relation
\be
{ \Omega_0 \over \rho_0^2 }U_{w0}O_s(\tau)U_{w0}^\dagger\mid_{\tau=\tau(t)}=O_0(t).
\ee
In Eq. (7), the overdot denotes the differentiation with respect to time $t$, 
while in Eq. (8) the notation $~"\mid_{\tau=\tau(t)}"$ is to mention that $\tau$ 
should be replaced by the function of $t$ satisfying the relation (6). 
In a different vein, the relation (8) has also been noticed in Ref. \cite{Seleznyova}.
Eqs. (5,8) imply the following relation in wave functions;
\be
\psi_n^0(x,t)=U_{w0}\psi_n^s|_{\tau=\tau(t)}.
\ee

As is well-known \cite{Shankar}, the simplest choice of $\{\psi_n^s|n=0,1,2,\cdots\}$ may 
be given as
\begin{eqnarray}
\psi_n^s(x,\tau)|_{\tau=\tau(t)}&=&
          {1\over \sqrt{2^nn!\sqrt{\pi\hbar}}} e^{-i(n+{1\over 2})\tau} \cr
&&    \times
  \exp[-{x^2\over 2\hbar}]H_n({1\over \sqrt{\hbar}}x)|_{\tau=\tau(t)}\\
&=&{1\over \sqrt{2^nn!\sqrt{\pi\hbar}}}
          ({u_0(t)-iv_0(t) \over \rho_0(t)})^{n+1/2}\cr
&&    \times\exp[-{x^2\over 2\hbar}+ic_0]H_n({1\over \sqrt{\hbar}}x),
\end{eqnarray} 
where $c_0$ is an arbitrary real number which will be set to zero from now on.
In obtaining Eq. (11), we make use of the fact:
\be
d\tau={\Omega_0 \over \rho_0^2}dt=i({\dot{u}_0 -i\dot{v}_0 \over u_0 -i v_0}
      -{\dot{\rho}_0\over \rho_0})dt.
\ee
In order to find a general expression of $\psi_n^0(x,t)$, we consider another
unitary transformation.
By defining $\delta_{u_1}(t)$ through the relations
\be
\dot{\delta}_{u_1}={1\over 2}w_0^2u_1^2-{1\over 2}\dot{u}_1^2
\ee
where $u_1$ is a linear combination of $u_0(t)$ and $v_0(t)$,
one may find that the unitary operator $U_f$ given as \cite{SongU}
\be
U_f=\exp[{i\over \hbar}(\dot{u}_1x+\delta_{u_1}(t)]\exp(-{i \over \hbar}u_1p)
\ee
satisfies the following relation
\be
U_fO_0U_f^\dagger=O_0.
\ee
Therefore, the wave functions $\psi_n^0$ satisfying Schr\"{o}dinger equation of
Eq. (4) may in general be written as
\begin{eqnarray}
\psi_n^0(x,t)&=&U_fU_{w0}\psi_n^s(x,\tau)\mid_{\tau=\tau(t)}\cr
&=&{1\over \sqrt{2^nn!\rho_0(t)}}({\Omega_0\over \pi\hbar})^{1/4}
({u_0(t)-iv_0(t) \over \rho_0(t)})^{n+1/2}\cr
&&\times \exp[{i\over \hbar}(\dot{u}_1(t)x+\delta_{u_1}(t))]\cr
&&\times 
\exp[{(x-u_1(t))^2\over 2\hbar}
(-{\Omega_0\over \rho_0^2(t)}+i{\dot{\rho}_0 \over \rho_0})]
\cr&&\times
H_n(\sqrt{\Omega_0\over \hbar}{x-u_1(t) \over \rho_0(t)}).
\end{eqnarray} 
This wave function, of course, agrees with the known one \cite{Song,SongU,BFL}
if we consider $u_1$ as a (fictitious) particular solution. 
If $u_1=0$, the wave function given in Eq. (16) also agrees with that in 
Refs. \cite{KL,Um,Ji,Ped2}. 

It may be interesting to find that how many free parameters are in the wave function
$\psi_n^0(x,t)$. First, there are two parameters in determining $u_1(t)$.
In the case of $u_1=0$, one may think that there are four parameters which come 
from determining $u_0(t),v_0(t)$. However, one of them is not a free parameter, 
since the wave functions are invariant under the multiplication of $u_0(t)$ and 
$v_0(t)$ with same constant factor. For the simple harmonic oscillator of  
time-translational symmetry, one of the remaining three parameters of $u_1=0$ is 
simply related to a time-shifting of the wave function. This can be seen from the fact
that, for the unit frequency case, the $u_0$ and $v_0$ can be taken as $\cos(t+t_0)$ and 
$C\sin(t+\beta+t_0)$, respectively, with real constants $C,\beta,t_0$.

If one considers $\rho_s(\tau)$ satisfying
\be
{d^2 \over d\tau^2}{\rho_s}+\rho_s - {\Omega_s^2 \over \rho_s^3}=0,
\ee 
and a simple harmonic oscillator of unit mass and frequency and with time 
$\tau'$ which is related to $\tau$ as 
\[ 
d\tau'={\Omega_s \over \rho_s^2}d\tau,
\]
by defining 
\be
U_s=\exp({i{d\rho_s/ d\tau} \over 2\hbar \rho_s}x^2)
  \exp[-{i\over 2\hbar}(\ln{\rho_s\over \sqrt{\Omega_s}})(xp+px)],
\ee
one may find that
\be
{\Omega_s \over \rho_s^2}U_sO_s(\tau')U_s^\dagger\mid_{\tau'=\tau'(\tau)}
=O_s(\tau).
\ee
The wave functions $\tilde{\psi}_n^s(\tau)$ defined by
\[
\tilde{\psi}_n^s(\tau) \equiv U_s\psi_n^s(x,\tau')\mid_{\tau'=\tau'(\tau)} 
\] 
then satisfy the Schr\"{o}dinger equation $O_s(\tau)\tilde{\psi}_n^s=0$; 
In fact, $\tilde{\psi}_n^s(\tau)$ is closely related to the wave functions of 
the squeezed states \cite{Stoler,Gilmore,Klauder}. 

One may think that a more general expression of the unitary operator, $U_{w0}$, may be 
obtained by combining use of $U_{w0}$ and $U_s$. This, however, is not the case 
as can be seen from the relation 
\be
U_{w0}(\rho_0,\Omega_0)U_s\mid_{\tau=\tau(t)}
=U_{w0}(\rho_0\rho_s,\Omega_0\Omega_s)\mid_{\tau=\tau(t)},
\ee
which is in accordance with the number counting of free parameters in 
$\psi_n^0(x,t)$.

The harmonic oscillator of unit mass and frequency with an inverse-square 
potential is described by the Hamiltonian \cite{Calogero}
\be
H_{in}^s={p^2 \over 2}+{x^2 \over 2}+{g \over x^2}.
\ee 
We only consider the case of $g>-\hbar^2/8$, and the region of $x>0$.
By defining $\alpha={1\over 2}(1+8g/\hbar^2)^{1/2}$ and  
\be
O_s^{in}(\tau)=-i\hbar{\partial \over \partial \tau} + H_{in}^s,
\ee 
the wave function $\phi_n^s$ satisfying the Schr\"{o}dinger equation 
$O_s^{in}(\tau)\phi_n^s=0$ is given as \cite{Calogero}
\begin{eqnarray}
\phi_n^s &\equiv& \langle x|\phi_n^s\rangle\cr
&=& ({4\over \hbar})^{1/4}({\Gamma(n+1) \over \Gamma(n+\alpha+1})^{1/2}
       e^{-i(2n+\alpha+1)\tau}\cr
&&\times   ({x^2 \over \hbar})^{(2\alpha+1)/4}  
   \exp(-{x^2 \over 2\hbar}) L_n^\alpha({x^2 \over \hbar}).
\end{eqnarray}
By defining $O_{0}^{in}$ as
\be 
O_{0}^{in}(t)= -i\hbar{\partial \over \partial t}
      +{p^2 \over 2}+w_0^2(t){x^2 \over 2}+{g \over x^2},
\ee
as in the case without the inverse-square term, one may find the relation
\be
{\Omega_0 \over \rho_0^2}U_{w0}O_s^{in}(\tau)U_{w0}^\dagger\mid_{\tau=\tau(t)}
   =O_0^{in}(t).
\ee
In deriving Eq. (25), we make use of the commutator relation
\be
[xp+px, {1\over x^2}]=4i\hbar {1\over x^2}.
\ee
For a further generalization, we define a unitary operator
\be
U_g=\exp[{i\over \hbar}(Max^2-{\dot{M} \over 4})x^2]
          \exp[{i{\ln M \over 4 \hbar}(xp+px)}],
\ee
where $M$ is a positive function of $t$, and $a(t)$ is a real function.
One may then easily find the relation
\begin{eqnarray}
U_gO_0^{in}U_g^\dagger&\equiv& O^{in}\\
&=&-i\hbar{\partial\over \partial t}+H_{in},
\end{eqnarray}
where (see Ref. \cite{Song}) 
\begin{eqnarray}
H_{in}&=&{p^2\over 2M(t)}-a(t)(xp+px) \cr
      & &+{1\over 2}M(t)c(t)x^2+{g\over M(t)}{1\over x^2} 
\end{eqnarray}
with
\[
c(t)=w_0^2(t)+{1\over \sqrt{M}}{d^2\sqrt{M} \over dt^2} +4a^2-2{1\over M}{d\over dt}(Ma).
\] 
For convenience \cite{SongU}, we consider the equation 
\be
{d\over dt}(M\dot{x}) +w^2(t)x=0, 
\ee
where $w^2(t)=w_0^2(t)+{1\over \sqrt{M}}{d^2\sqrt{M} \over dt^2}.$
The two linearly independent solutions $u(t),v(t)$ of Eq. (31) can be given
from $u_0(t),v_0(t)$ as $u(t)={u_0 \over \sqrt{M}}, v(t)={v_0 \over \sqrt{M}}$,
so that one may find the relation $\Omega_0=M(\dot{v}u-\dot{u}v)$.
We also define the $\rho(t)$ as $\rho(t)={\rho_0 \over \sqrt{M}}$.
The wave function $\phi_n$ of the system described by the Hamiltonian 
$H_{in}(x,p,t)$ can then be obtained as
\begin{eqnarray}
\phi_n&=&U_G\phi_n^s(\tau)\mid_{\tau=\tau(t)}\\
  &=&({4\Omega_0\over \hbar\rho^2})^{1/4}({\Gamma(n+1) \over \Gamma(n+\alpha+1})^{1/2}\cr
&&\times ({u-iv\over \rho})^{(2n+\alpha+1)}  
     ({\Omega_0 x^2 \over \hbar\rho^2})^{(2\alpha+1)/4}    \cr
&&\times \exp[-{x^2 \over 2\hbar}({\Omega_0 \over \rho^2}-iM{\dot{\rho}\over \rho}-2iMa)]
 L_n^\alpha({\Omega_0 x^2 \over \hbar\rho^2}),
\end{eqnarray}
where
\be
U_G=U_gU_{w0}.
\ee
For $a=0$, the wave functions $\phi_n$ agree with those in Refs. \cite{KL,Ped2}. 
As in Ref. \cite{Song}, by considering the kernel of the
system \cite{KL}, it may be easy to see that the wave functions $\phi_n(x,t)$ 
form a complete set. The form of $\phi_n$ in Eq. (33) indicates that, even for the 
system described by the constant Hamiltonian $H_{in}^s$ given in Eq. (21),
there are wave functions whose probability density distributions pulsate as in
those of the squeezed states. 

For the system of the Hamiltonian $H_{in}$, if $M(t),w_0^2(t),$ and $a(t)$ are periodic 
with a period $T$, one may study the non-adiabatic geometric phases \cite{AA,SW}.
The wave function $\phi_n$ is (quasi)periodic, only if $\rho(t)$ is periodic. 
The condition for periodic $\rho(t)$ with the period $T'$ $(=T$ or $2T)$
has been analyzed in Ref. \cite{SongG}. Here, we only consider the case of such a
periodic $\rho(t)$. The overall phase change of $\phi_n$
under the $T'$ evolution is given as 
\[
\chi_n=-(2n+\alpha+1)\int_0^{T'}{\Omega_0 \over\rho_0^2(t)}dt.
\] 
The expectation value of the $H_{in}$ can be evaluated 
by making use of the relation
\be
H_{in}\phi_n=(i\hbar{\partial\over \partial t}U^G)\phi_n^s
+U_Gi\hbar{\partial \over \partial t}\phi_n^s.
\ee
From the fact that 
\be
i\hbar{\partial \over \partial t}\phi_n^s=
i\hbar{d\tau\over dt}{\partial \over \partial \tau}\phi_n^s=
(2n+\alpha+1)\hbar{\Omega_0 \over\rho_0^2(t)}\phi_n^s,
\ee
one may find that the geometric phase $\gamma_n$ for the wave function $\phi_n$ 
under the $T'$ evolution is written as
\begin{eqnarray}
\gamma_n&=&\chi_n+{1\over \hbar}\int_0^{T'} \langle \phi_n|H_{in}|\phi_n \rangle dt\cr
&=&{1\over \hbar\Omega_0}\int_0^{T'}(M\dot{\rho}^2+2Ma\rho\dot{\rho})dt
   \langle \phi_n^s|x^2|\phi_n^s\rangle\cr
&=&(2n+\alpha+1){1\over \Omega_0}\int_0^{T'}(M\dot{\rho}^2+2Ma\rho\dot{\rho})dt.
\end{eqnarray}

The unitary operators can be used in finding the {\em exact} invariants
for the cases without and with the inverse-square potential from $H^s$ and $H_{in}^s$, 
respectively. First of all, it is clear that $H^s$ and $H_{in}^s$ are invariants 
in the systems they describe, respectively. For the system described by $H_0(x,p,t)$, 
if we only consider the case of $u_1=0$, the invariant $I_0$ is obtained by applying
the unitary transformation to the invariant $H^s$
\begin{eqnarray}
I_0&=&U_{w0}H^sU_{w0}^\dagger\cr
   &=&{1\over 2\Omega_0}[({\Omega_0x\over \rho_0})^2+(\rho_0p-\dot{\rho}x)^2],
\end{eqnarray}
which agrees with those in Refs. \cite{Lewis,LR,RR,KL,Um,Ji}.
For the system described by $H_{in}(x,p,t)$, the invariant is again given from the
invariant $H_{in}^s$ as
\begin{eqnarray}
I_{in}&=&U_GH_{in}^sU_G^\dagger\cr
   &=&{1\over 2\Omega_0}[({\Omega_0x\over \rho})^2\cr
   & &~~~~~+\{\rho p-(M\dot{\rho}+2Ma\rho)x\}^2+2\rho^2{g\over x^2}].
\end{eqnarray}
For the case of $a=0$, the invariant $I_{in}$ reduces to the known one 
\cite{Ped2}. One can explicitly check that the invariant $I_{in}$ indeed satisfies 
the relation
\be 
i\hbar{\partial I_{in} \over \partial t} +[I_{in},H_{in}]=0.
\ee
Alternatively, making uses of Eqs. (35,36) and  relying on the completeness of 
the set $\{\phi_n^s|~n=0,1,2,\cdots\}$, a simple proof of Eq. (40) may also be 
possible.

In summary, we have found a unitary operator which transforms a harmonic oscillator 
system of time-dependent frequency into that of a simple harmonic oscillator of 
different time-scale, with and without the inverse-square potential. Making use
of the unitary operator, the exact invariants and wave functions of the 
time-dependent systems have been evaluated from the well-known results in the 
corresponding system of constant Hamiltonians.
It should be mentioned, however, that the classical solutions of the time-dependent 
harmonic oscillator system must be found for actual applications, 
while the classical equation (see Eq. (2)) is formally equivalent to 
a one-dimensional time-independent Schr\"{o}dinger equation 
(of arbitrary potential).
The classical correspondent of unitary transformation is the 
canonical transformation which has been studied in the model 
\cite{BFL,LL}. It would be interesting if the relationship  
could be used in finding relations among the quantities in classical 
and quantum mechanics such as that between the geometric phases and 
the Hannay's angle (see Ref. \cite{SongG}).

\end{multicols}

\end{document}